# The Metric of Quantum States Revisited

## Aalok Pandya† and Ashok K. Nagawat

Department of Physics, University of Rajasthan, Jaipur 302004, India

**Abstract**

A generalised definition of the metric of quantum states is proposed by using the techniques of differential geometry. The metric of quantum state space derived earlier by Anandan, is reproduced and verified here by this generalised definition. The metric of quantum states in the configuration space and its possible geometrical framework is explored. Also, invariance of the metric of quantum states under local gauge transformations, coordinate transformations, and the relativistic transformations is discussed.

___________________________

† Also at Jaipur Engineering College and Research Centre (JECRC), Jaipur 303905, India.

*E-Mail*: aalok@uniraj.ernet.in

# 1. INTRODUCTION

 In recent years as the study of geometry of the quantum state space has gained prominence, it has inspired many to explore further geometrical features in quantum mechanics. The metric of quantum state space, introduced by J. Anandan [1], and Provost and Vallee [2], is the source of motivation for the present work. Anandan followed up this prescription of metric by yet another derivation [3] to describe the metric of quantum state space with the specific form of metric coefficients. However, the metric tensor of quantum state space, was defined for the first time by Provost and Vallee [2] from underlying Hilbert space structure for any sub-manifold of quantum states by calculating the distance function between two quantum states. Since then, many attempts have been made to explain and to analyse metric of quantum evolution in the projective Hilbert space [1-9] $\mathscr{P}$. Recently, researchers studying gravity have also shown considerable interest in the geometric structures in quantum mechanics in general and projective Hilbert space $\mathscr{P}$ in specific [4-7].

The invariant $ds$ of the metric defined by Anandan and others as the distance between two quantum states $|\Psi(t)\rangle$ and $|\Psi(t+dt)\rangle$, is of the form:

$$ds^2 = \frac{4}{\hbar^2}\left(\langle\Psi|H^2|\Psi\rangle - \langle\Psi|H|\Psi\rangle^2\right)dt^2, \qquad (1)$$

and $ds^2 = 4(1 - |\langle\Psi(t)|\Psi(t+dt)\rangle|^2)$. $\qquad (2)$

This is also known as Fubini Study metric of the Ray Space. Here, $H$ is Hamiltonian and the invariant $ds$ can be regarded as the distance between points $p$ and $p'$ in the projective Hilbert space $\mathscr{P}$. And, $|\Psi(t)\rangle$ and $|\Psi(t+dt)\rangle$, are two normalised states contained in $p$ and $p'$; clearly with the condition $s(p,p') \geq 0$. This metric when restricted to P$_1$(𝒞), is to be regarded as a 2-sphere with radius embedded in a real three-dimensional Euclidean space, and $s(p,p')$ is then the straight line, or better called geodesic distance between $p$ and $p'$ on this sphere [3]. Alternatively, Anandan [3] has formulated this metric as:

$$ds^2 = 2\left(\left\langle\frac{\partial\tilde{\Psi}}{\partial z^\mu}\bigg|\frac{\partial\tilde{\Psi}}{\partial z^\nu}\right\rangle - \left\langle\frac{\partial\tilde{\Psi}}{\partial z^\mu}\bigg|\tilde{\Psi}\right\rangle\left\langle\tilde{\Psi}\bigg|\frac{\partial\tilde{\Psi}}{\partial z^\nu}\right\rangle\right)d\bar{z}^\mu dz^\nu = 2g_{\bar{\mu}\nu}d\bar{z}^\mu dz^\nu, \tag{3}$$

for the complex coordinates $Z^\mu$ in 𝒫. This can be regarded as an alternative definition of the Fubini-Study metric, valid for an infinite dimensional 𝓗.

This metric is real and positive definite [3]. We cannot expect a metric with the signature of Minkowski space in the study of the metric of quantum state space, as the metric of quantum state space is in the projective Hilbert space and therefore it is always positive definite. However, we can define the metric of quantum states in the configuration space, but such a metric need not be always positive definite. To be precise, the metric of quantum state space is a metric on the underlying manifold which the quantum states form or belong to, and therefore, it is different from the metric of space-time or any other metric associated with the quantum states.

A quantum state in the Hilbert space corresponds to a point in the projective Hilbert space, by means of projections. And two points in the projective Hilbert space can lie on a line which stands for neighborhood in topological sense provided the corresponding two states in the Hilbert space are connected by means of invariance under local gauge transformations. The basic objective of Anandan, Provost, Vallee and others, behind formulation of the metric of quantum state space, was to seek invariance in the quantum evolution under the local gauge transformations [1-3, 8-13]. One can verify this fact from the equation (3); where, there are two parts in the expression of metric coefficient $g_{\mu\nu}$, such that whenever the first part picks up an additional term due to local gauge transformation, it gets cancelled by a similar extra term picked up by the second part. Thus, the metric of quantum state space is invariant under the local gauge transformations in addition to the invariance under coordinate transformations. As rightly pointed out by Minic and Tze, everything we know about quantum mechanics is in fact contained in the geometry [4-6] of $CP(N)$. Entanglements come from the embeddings of the products of two complex projective spaces in a higher dimensional one; geometric phase stem from the symplectic structure of $CP(N)$, quantum logic, algebraic approaches to quantum mechanics etc, are all contained in the geometric and symplectic structure of complex projective spaces [4-6]. While we only consider here the finite dimensional case, the same geometric approach is extendible to generic infinite dimensional quantum mechanical systems, including field theory. Finally, the following three lemmas summarize this discussion as:

(i) The Fubini-Study metric as given in the equation (2) and (3) in the limit $\hbar \to 0$ becomes a spatial metric, provided the configuration space for the quantum system under consideration is space-time. For example, if we consider a particle moving in 3-dimensional Euclidean space, then the quantum metric for the Gaussian coherent state $\Psi_l(x) \sim \exp\left(-\frac{(\vec{x}-\vec{l})^2}{\delta l^2}\right)$ yields the natural metric in the configuration space, in the limit

$$\hbar \to 0 \text{, becomes } ds^2 = \frac{d\vec{l}^2}{\delta l^2}. \tag{4}$$

(ii) Similarly, the time parameter of the evolution equation can be related to the quantum metric *via*

$$\hbar ds = \Delta E dt, \quad \Delta E^2 \equiv \langle \Psi | H^2 | \Psi \rangle - \langle \Psi | H | \Psi \rangle^2. \tag{5}$$

(iii) Finally, the Schrödinger equation can be viewed as a geodesic equation on a $CP(N) = \frac{U(N+1)}{U(N) \times U(1)}$ as:

$$\frac{du^a}{ds} + \Gamma^a_{bc} u^b u^c = \frac{1}{2\Delta E} Tr(HF^a_b) u^b. \tag{6}$$

Here $u^a = \frac{dz^a}{ds}$ where $z^a$ denote the complex coordinates on $CP(N)$, $\Gamma^a_{bc}$ is the connection obtained from the Fubini-Study metric, and $F_{ab}$ is the canonical curvature 2-form valued in the holonomy gauge group $U(N) \times U(1)$. Here, Hilbert space is $N+1$ dimensional and the projective Hilbert space has dimensions $N$.

The motivation behind our exercise in this paper is two fold: firstly, to propose generalised definition of the metric of quantum states by using techniques of differential geometry. And secondly, to think beyond the quantum state space by exploring the prospects of metric of quantum states in the configuration space. We also discuss the prerequisite geometrical framework of any possible metric on space-time manifold with pseudo-Riemannian structures.

## 2. GENERALISED DEFINITION OF THE METRIC OF QUANTUM STATES

The definition of the total covariant derivative of any function implies:

$$\nabla \Psi = \nabla_{\partial/\partial x_\mu} \Psi \otimes dx^\mu. \tag{7}$$

Since the exterior derivative of a (total) differential form vanishes as

$$d(d\Psi) = 0. \tag{8}$$

Now, taking product of the equation $\nabla \Psi = \nabla_\mu \Psi \otimes dx^\mu$, with equation

$$\nabla \Psi = \nabla_\nu \Psi \otimes dx^\nu, \tag{9}$$

we define invariant: $ds^2 = |\nabla \Psi|^2 = (\nabla_\mu \Psi)(\nabla_\nu \Psi) dx^\mu dx^\nu.$ \hfill (10)

The metric tensor $g_{\mu\nu}$ for the above can be given as:

$$g_{\mu\nu} = \text{Re}[(\nabla_\mu \Psi)(\nabla_\nu \Psi)]. \tag{11}$$

Alternatively, one can also write the symmetric tensor $g_{\mu\nu}$ as

$$g_{\mu\nu} = (\nabla_\mu \Psi)^*(\nabla_\nu \Psi) = \frac{1}{2}\left[(\nabla_\mu \Psi)^*(\nabla_\nu \Psi) + ((\nabla_\mu \Psi)^*(\nabla_\nu \Psi))^*\right]$$

$$= \frac{1}{2}\left[(\nabla_\mu \Psi)^*(\nabla_\nu \Psi) + (\nabla_\nu \Psi)^*(\nabla_\mu \Psi)\right]. \tag{12}$$

We find that this generalised definition satisfies all geometrical requirements of metric structure. The $g_{\mu\nu}$ can be transformed under co-ordinate transformations and therefore it is a tensor of second rank. The quantity $g_{\mu\nu}$, constitutes a real (or hermitian) matrix of order $n \times n$, for $\mu = 1,...,n$ and $\nu = 1,..,n$. However, the indices $\mu$ and $\nu$ vary from 1 to…4, when the metric is in the configuration space of space- time. We discuss the invariance and the possible geometric framework for prospective metric structures in the following discussions.

Following this generalised definition, we deduce the metric of quantum state space, and the metric of quantum states in the configuration space. We also illustrate several examples on it.

## 3. THE METRIC OF QUANTUM STATE SPACE

From the generalized definition discussed here, we reproduce the expression of the metric of quantum state space. We consider a quantum state $\Psi \equiv \Psi\{\lambda\}$, $\forall \Psi \in \mathcal{H}$, and the corresponding covariant derivative for the quantum states [3] is given by:

$$\nabla^\lambda \Psi \equiv \left|\frac{d\Psi}{d\lambda}\right\rangle + \left\langle\Psi\left|\frac{d\Psi}{d\lambda}\right\rangle|\Psi\rangle. \tag{13}$$

Here, $\lambda$ in equation (13) can be any local co-ordinate on $\mathcal{P}$. Applying this covariant derivative to the definition of metric in eq. (12) we obtain the desired metric coefficients:

$$g_{\lambda\lambda} = \left[(\nabla_\lambda \Psi)^*(\nabla_\lambda \Psi)\right] = \frac{1}{2}\left[(\nabla_\lambda \Psi)^*(\nabla_\lambda \Psi) + \left((\nabla_\lambda \Psi)^*(\nabla_\lambda \Psi)\right)^*\right]$$

$$= \left[\left(\left\langle\frac{\partial\Psi}{\partial\lambda}\right| - \langle\Psi|\left\langle\frac{\partial\Psi}{\partial\lambda}\right|\Psi\rangle\right)\left(\left|\frac{\partial\Psi}{\partial\lambda}\right\rangle - \left\langle\Psi\left|\frac{\partial\Psi}{\partial\lambda}\right\rangle|\Psi\rangle\right)\right],$$

$$= \left\langle\frac{\partial\Psi}{\partial\lambda}\bigg|\frac{\partial\Psi}{\partial\lambda}\right\rangle - \left\langle\frac{\partial\Psi}{\partial\lambda}\bigg|\Psi\right\rangle\left\langle\Psi\bigg|\frac{\partial\Psi}{\partial\lambda}\right\rangle - \left\langle\frac{\partial\Psi}{\partial\lambda}\bigg|\Psi\right\rangle\left\langle\Psi\bigg|\frac{\partial\Psi}{\partial\lambda}\right\rangle + \left\langle\frac{\partial\Psi}{\partial\lambda}\bigg|\Psi\right\rangle\left\langle\Psi\bigg|\frac{\partial\Psi}{\partial\lambda}\right\rangle\langle\Psi|\Psi\rangle.$$

Which gives $g_{\lambda\lambda} = \left[\left\langle\frac{\partial\Psi}{\partial\lambda}\bigg|\frac{\partial\Psi}{\partial\lambda}\right\rangle - \left\langle\frac{\partial\Psi}{\partial\lambda}\bigg|\Psi\right\rangle\left\langle\Psi\bigg|\frac{\partial\Psi}{\partial\lambda}\right\rangle\right].$  (14)

Also, we can write it in a generalized way as:

$$g_{\bar{\mu}\nu} = \left[\left\langle\frac{\partial\Psi}{\partial x_\mu}\bigg|\frac{\partial\Psi}{\partial x_\nu}\right\rangle - \left\langle\frac{\partial\Psi}{\partial x_\mu}\bigg|\Psi\right\rangle\left\langle\Psi\bigg|\frac{\partial\Psi}{\partial x_\nu}\right\rangle\right]. \tag{15}$$

This is same as the metric of quantum state space, formulated by Anandan and others [1, 3-5, 8-11] for the real local coordinates $x^\mu$. But this metric is no more on Kähler manifold. If the metric of quantum states is defined with local co-ordinates that are not complex, it lies on the base manifold with Riemannian character, and the local gauge group $GL(n,R)$ is also admissible, where $n$ is the dimensionality of the space-time.

The generalized formulation discussed here, in turn verifies and validates Anandan's formulation of the metric of quantum state space, which was derived by Taylor's expansion and other specific methods.

*Further examination of the metric of quantum state space*

We now examine the metric of quantum state space by yet another exercise. This is illustration of the fact that the application of covariant derivative makes quantum evolution independent of the type of quantum evolution, be it relativistic or non-relativistic. We consider the Klein-Gordon equation as follow:

$$-\nabla^\mu \nabla_\mu \Psi = \frac{m_0^2 c^2}{\hbar^2} \Psi. \tag{16}$$

Multiplying it from left by the hermitian conjugate of $\Psi$, we get

$$-\Psi^* \nabla^\mu \nabla_\mu \Psi = \frac{m_0^2 c^2}{\hbar^2} \Psi^* \Psi. \tag{17}$$

This expression is covariant and also invariant under local gauge transformations. Being inspired by the covariance and the invariance of this expression, we formulate a metric with the help of it. For $\Psi \equiv \Psi(x^\mu)$, we can write

$$ds^2 = \Psi^* \nabla_\mu \nabla_\nu \Psi \, dx^\mu dx^\nu, \tag{18}$$

so that $g_{\mu\nu} = \Psi^* \nabla_\mu \nabla_\nu \Psi = \frac{1}{2}\left[(\Psi^* \nabla_\mu \nabla_\nu \Psi) + (\Psi^* \nabla_\mu \nabla_\nu \Psi)^*\right]$. (19)

We notice that this definition of $g_{\mu\nu}$ involves a second order derivative and ordinary second order partial derivative does not satisfy transformation properties. Therefore, it becomes inevitable to apply here the covariant form of derivative defined in eq. (13). The metric coefficient of the above invariant thus takes the following form:

$$g_{\mu\nu} = \left\langle \Psi \left| \left[ \frac{d}{dx_\mu} \left( \left| \frac{d\Psi}{dx_\nu} \right\rangle - \left\langle \Psi \left| \frac{d\Psi}{dx_\nu} \right\rangle | \Psi \right\rangle \right) - \left\langle \Psi \left| \frac{d\Psi}{dx_\mu} \right\rangle \left( \left| \frac{d\Psi}{dx_\nu} \right\rangle - \left\langle \Psi \left| \frac{d\Psi}{dx_\nu} \right\rangle | \Psi \right\rangle \right) \right] \right. \right\rangle,$$

or $g_{\mu\nu} = -\left[ \left\langle \frac{\partial \Psi}{\partial x_\mu} \middle| \frac{\partial \Psi}{\partial x_\nu} \right\rangle - \left\langle \frac{\partial \Psi}{\partial x_\mu} \middle| \Psi \right\rangle \left\langle \Psi \middle| \frac{\partial \Psi}{\partial x_\nu} \right\rangle \right]$. (20)

Multiplying it by negative sign, we obtain the same metric of quantum state space as described earlier in equation (14) and in references: (1-3). Thus, we again confirm the unique form of the invariant expression of the metric of quantum state space. It should be noticed that the normalization in relativistic quantum mechanics is given by the criteria prescribed by Weinberg [14].

## 4. THE METRIC OF QUANTUM STATES IN THE CONFIGURATION SPACE

In a generalised formalism of geometric quantum mechanics, coordinates are not meaningful. On a Kahler manifold in the quantum state space, invariance under the local gauge transformations is same as invariance under the coordinate transformations. This is with the reason that in the quantum state space, quantum states themselves could play the role of coordinates. The definition of the metric tensor in (11) and (12) involves only first order derivatives, thus even if we use ordinary partial derivatives instead of the covariant derivative defined in (13), the metric properties of $g_{\mu\nu}$ remain unaffected. Also, even if we do not apply the complex conjugation, and consider only the real part of eq. (11), we still retain the metric structure. However, for such a metric positive-definiteness is no more assured, as it is not the metric of quantum state space. We redefine our metric as:

$$ds^2 = \text{Re }(\nabla\Psi)^2 = \text{Real Part}\left[(\nabla_\mu\Psi)(\nabla_\nu\Psi)\right]dx^\mu dx^\nu, \tag{21}$$

such that, $g_{\mu\nu} = \text{Real Part}\left[(\nabla_\mu\Psi)(\nabla_\nu\Psi)\right] = \text{Real Part}\left[\left(\dfrac{\partial\Psi}{\partial x_\mu}\right)\left(\dfrac{\partial\Psi}{\partial x_\nu}\right)\right].$  (22)

We observe that:

(i) The $ds$ being differential form guarantees invariance of this metric under the coordinate transformations.

(ii) And the quantity $g_{\mu\nu} = \left(\dfrac{\partial\Psi}{\partial x_\mu}\right)\left(\dfrac{\partial\Psi}{\partial x_\nu}\right)$ is a transformable quantity.

Since this is metric in the configuration space, the nature and signature of the metric will depend upon the choice of wave function. And a metric in the configuration space could be a metric on the space-time, wherever the configuration space coincides with the space-time [4-5]. However, anything more could be elaborated on this aspect, only on the further specification of a physical scenario. Thus, we cannot comment anything more here on this issue except its wider geometrical framework.

## 5. DISCUSSION

One may surprise, "How do we get different metric structures from a generalised definition?" Answer is simple! The coordinates used in case of metric of quantum states in ray space, are the local coordinates on the manifold of the quantum states in the projective Hilbert space $\mathscr{P}$. Where as, in case of metric in configuration space, the coordinates used are the coordinates in configuration space. Also, one could notice the reasons for invariance of the metric of quantum states in ray space under local gauge transformations. The 'connection'- $\left\langle\Psi\left|\dfrac{\partial\Psi}{\partial\lambda}\right.\right\rangle$ sitting inside the covariant derivative $\nabla^\lambda \Psi \equiv \left| \frac{d\Psi}{d\lambda} \right\rangle + \left\langle \Psi \left| \frac{d\Psi}{d\lambda} \right. \right\rangle |\Psi\rangle$, and having rooted its feet in local coordinates, always keeps connecting the initial state with the final state. This results into the invariance of the metric of the ray space under local gauge transformations, which is precisely the essence of the metric formulation in ray space. In case of metric in the configuration space it does not happen, and metric remains invariant only under coordinate transformations. It should be noticed that if the metric of quantum states is defined in the configuration space with the space-time co-ordinates, the base manifold **M** on which it lies, can carry a (pseudo) Riemannian metric as well, and the tetrad can naturally be chosen to bring the metric $g_{\mu\nu}$ to a diagonal Minkowski form, and then the Lorentz group $SO(3,1)$ appears as a local gauge group.

We clarify that the description of the metric of quantum states in the configuration space would not be merely for the sake of just another parameterization. Apart from the fundamental difference that, the metric of quantum state space is metric in the ray space and the metric otherwise stated is in the configuration space, there are many other differences:

(i) The signature of the metric of quantum state space is always positive definite. Where as, the signature of a possible metric in the configuration space need not be positive definite.

(ii) So far, we have encountered metric structures on three different manifolds: Kähler manifold or $CP(N)$, Riemannian manifold, and space-time (pseudo- Riemannian) manifold.

If the metric of quantum states is defined with the complex coordinates in the quantum state space, known as Fubini- Study metric, it lies on the Kähler manifold or $CP(N)$, which is identified with the quotient set $\dfrac{U(N+1)}{U(N)\times U(1)}$.

And if the metric of quantum states is defined with local co-ordinates that are not complex, it lies on the base manifold with Riemannian character, and the local gauge group $GL(n,R)$ is also admissible.

Whereas, if the metric of quantum states is defined in the configuration space with the space-time co-ordinates, the base manifold **M** on which it lies, carries a (pseudo) Riemannian metric as well, and the tetrad can naturally be chosen to bring the metric $g_{\mu\nu}$ to a diagonal Minkowski form. And then the Lorentz group $SO(3,1)$ could also appear as a local gauge group.

We must notice that the group symmetry observed in the quotient set $\dfrac{U(N+1)}{U(N)\times U(1)}$ in case of Fubini-Study metric is the symmetry over the transformations of the wave functions. Whereas, the group symmetry mentioned in the later cases as $GL(n,R)$ and $SO(3,1)$, if observed, can be due to the transformations of co-ordinates.

(iii) The metric of quantum state space is invariant under coordinate transformations as well as local gauge transformations. Where as, the metric in the configuration space need not be invariant under the local gauge transformations. But the metric in the configuration space is invariant at least under the coordinate transformations. Also, if the wave function subject to condition is relativistic, the metric could be invariant under the Lorentz' (relativistic) transformation as well.

(iv) Metric coefficients $g_{\mu\nu} = \left[ \left\langle \frac{\partial \Psi}{\partial x^{\mu}} \middle| \frac{\partial \Psi}{\partial x^{\nu}} \right\rangle - \left\langle \frac{\partial \Psi}{\partial x^{\mu}} \middle| \Psi \right\rangle \left\langle \Psi \middle| \frac{\partial \Psi}{\partial x^{\nu}} \right\rangle \right]$, defined in the metric of quantum state space, are under the integrals and therefore constant. Where as, the metric coefficients in the case of metric in the configuration space need not be constant.

(v) Since, the metric coefficients in the metric of quantum state space are constant, all their derivatives readily vanish. Where as, for the metric of quantum states in the configuration space, there is possibility that one can explore the other geometric features associated with a quantum state and the metric associated with it.

## ACKNOWLEDGEMENT

The authors wish to express their gratefulness to Prof. A. Ashtekar for his encouragement and suggestions to improve the manuscript. The authors also wish to express their gratefulness to late J. Anandan for the inputs he provided while working out this paper.

___

### Résumé

Une définition générale de la métrique des états quantiques a été calculée en utilisant les méthodes de géométrie différentielle. La métrique des états quantique calculée précédemment par Anandan est vérifiée et reproduite par cette définition générale. La métrique des états quantiques dans l'espace des configurations et sa possible représentation géométrique sont explorées. L'invariance de la métrique des états quantiques avec une transformation de jauge locale et un changement de coordonnées, ainsi que les transformations relativistes, sont également discutées.

___